\documentclass[fleqn,usenatbib]{mnras}

\usepackage{newtxtext,newtxmath}

\usepackage[T1]{fontenc}
\usepackage{ae,aecompl}

\usepackage{graphicx}	
\usepackage{amsmath}	
\usepackage{times}

\newcommand{\Mpc}{\ensuremath{~\mathrm{Mpc}}}
\newcommand{\hMpc}{\ensuremath{~h^{-1}\mathrm{Mpc}}}

\renewcommand{\arcmin}{\ensuremath{~\mathrm{arcmin}}}
\newcommand{\degSq}{\ensuremath{~\mathrm{deg}^2}}
\newcommand{\map}[1]{\ensuremath{#1 \times #1 \degSq}}

\newcommand{\cosmoslics}{\emph{cosmo}-SLICS}
\newcommand{\LSST}{\textsc{lsst}}
\newcommand{\Euclid}{\textsc{euclid}}

\title[Cosmology with WL voids]{Constraining cosmology with weak lensing voids}

\author[Davies et. al] 
{Christopher T. Davies,$^{1}$\thanks{E-mail:christopher.t.davies@durham.ac.uk (CTD)}
Marius Cautun,$^{2}$
Benjamin Giblin,$^{3}$
Baojiu Li,$^{1}$\newauthor
Joachim Harnois-Déraps,$^{3,4}$
and Yan-Chuan Cai$^{3}$
\\
$^{1}$Institute for Computational Cosmology, Department of Physics, Durham University, South Road, Durham DH1 3LE, UK\\
$^{2}$Leiden Observatory, Leiden University, PO Box 9513, NL-2300 RA Leiden, the Netherlands\\
$^{3}$Scottish Universities Physics Alliance, Institute for Astronomy, University of Edinburgh, Blackford Hill, Scotland, UK\\
$^4$Astrophysics Research Institute, Liverpool John Moores University, 146 Brownlow Hill, Liverpool, L3 5RF, UK\\
}

\date{Accepted XXX. Received YYY; in original form ZZZ}

\pubyear{2020}

\begin{document}
\label{firstpage}
\pagerange{\pageref{firstpage}--\pageref{lastpage}}
\maketitle

\begin{abstract}
Upcoming surveys such as \LSST{} and \Euclid{} will significantly improve the power of weak lensing as a cosmological probe. To maximise the information that can be extracted from these surveys, it is important to explore novel statistics that complement standard weak lensing statistics such as the shear-shear correlation function and peak counts. In this work, we use a recently proposed weak lensing observable --- weak lensing voids --- to make parameter constraint forecasts for an $\LSST$-like survey. We use the \cosmoslics{} $w$CDM simulation suite to measure void statistics as a function of cosmological parameters. The simulation data is used to train a Gaussian process regression emulator that we use to generate likelihood contours and provide parameter constraints from mock observations. We find that the void abundance is more constraining than the tangential shear profiles, though the combination of the two gives additional constraining power. We forecast that without tomographic decomposition, these void statistics can constrain the matter fluctuation amplitude, $S_8$ within 0.3\% (68\% confidence interval), while offering 1.5, 1.5 and 2.7\% precision on the matter density parameter, $\Omega_{\rm m}$, the reduced Hubble constant, $h$, and the dark energy equation of state parameter, $w_0$, respectively. These results are tighter than the constraints from the shear-shear correlation function with the same observational specifications for $\Omega_m$, $S_8$ and $w_0$. The constraints from the WL voids also have complementary parameter degeneracy directions to the shear 2PCF for all combinations of parameters that include $h$, making weak lensing void statistics a promising cosmological probe. 
\end{abstract}

\begin{keywords}
gravitational lensing: weak -- large-scale structure of universe -- cosmology: theory -- methods: data analysis
\end{keywords}



\section{Introduction}

The standard model of cosmology, in which the dominant matter component consists of cold dark matter (CDM) while the late-time accelerated expansion is driven by a positive cosmological constant, $\Lambda$, is highly successful at describing a number of independent observations, which constrain these parameters with a large degree of concordance. Notably, measurement of fluctuations in the Cosmic Microwave Background (CMB) \citep{Planck2018}, provides measurements of the present day expansion rate of the universe $H_{0}$, the matter density parameter $\Omega_{\rm m}$ and the matter fluctuation amplitude $\sigma_8$. 

Another promising observational probe that is sensitive to and can be used to constrain many cosmological parameters is gravitational lensing, a phenomenon according to which the light of distant source images is distorted by the gravitational potentials of the foreground matter. In the strong lensing regime distant galaxies are visibly distorted into large arcs. In the weak lensing (WL) regime, which is the focus of this study, this effect is much smaller, and the weak lensing signal is measured through the correlations in distortions of many source galaxies \citep{Bacon2000,Kaiser2000,VanWaerbeke2000,Wittman2000}. This allows us to probe the total matter distribution of the Universe on the largest scales \citep[see][for reviews]{Bartelmann2001,Kilbinger2015}, and offers a powerful method to study the clustering of dark matter and its evolution.

Some of the most recent WL observations that supplement the parameter measurements from the CMB include the DES \citep{Troxel2018} \footnote{\url{https://www.darkenergysurvey.org/}}, HSC \citep{Hikage2019} \footnote{\url{https://hsc.mtk.nao.ac.jp/ssp/}} and KiDS \citep{Asgari2020} \footnote{\url{http://kids.strw.leidenuniv.nl/}} WL surveys. However, all of these surveys measure lower values of $\sigma_8$ compared to Planck, with a statistically significant disagreement arising in the comparison between the Planck and KiDS constraints. This is one example of the parameter tensions that have arisen in recent years, where different observations point to slightly different values of certain cosmological parameters, implying the presence of either unaccounted for systematics or new physics which are unaccounted for. Another example is the $H_0$ tension, where multiple observations find that measurements from the early Universe are broadly inconsistent with those of the late universe \citep{Verde2019}, particularly the distance scale measurement of $H_0$ based on Cepheids by the SH0ES collaboration \citep{Riess:2019cxk}. 

In order to address these parameter tensions, it is important to measure cosmological parameters as precisely as possible, by maximising the information that can be extracted from a given survey. The standard approach for weak lensing surveys is to measure $\Lambda\rm{CDM}$ parameters with two-point statistics such as the shear-shear correlation function or the convergence power spectrum  \citep{Schneider2002,Semboloni2006,Hoekstra2006,Fu2008,Heymans2012,Kilbinger2013,Hildebrandt2017,Troxel2018,Hikage2019,Aihara2019,Asgari2020}. However, two-point statistics do not capture non-Gaussian information, and weak lensing data are highly non-Gaussian due to the non-linear evolution of the Universe. To address this loss, many complimentary statistics have been developed, which encapsulate information beyond two-point statistics. A common and popular example is the abundance of WL peaks, which has been shown to be complimentary to the two-point function and helps break the $\Omega_m$-$\sigma_8$ parameter degeneracy \citep{Jain2000,Pen2003,Dietrich2010}. Peaks are also shown to outperform the standard methods for constraining the sum of neutrino mass \citep{Li2018} and $w_0$ \cite{Martinet2020}. By including complimentary statistics, the measurement errors on cosmological parameters can be reduced, which will help inform the statistical significance of any parameter tensions between multiple observations. 

The goal of this paper is to present parameter constraint forecasts for one such complimentary probe, WL voids. Voids are typically identified within the full 3D distribution of matter as regions of low matter density or low tracer density, for which void statistics such as their abundance, radial profiles and shapes contain useful non-Gaussian information \citep[see, e.g.,][]{White1979,Fry1986,Lee2009,Biswas2010,Bos2012,Lavaux2012,Jennings2013,Hamaus2014}. Most studies use galaxy voids, which are identified as underdense regions in the galaxy distribution \citep[e.g.,][]{Pan2012,Paz2013,Sutter2014,Cautun2016,Nadathur2016,Mao2017,Pollina2019,Hamaus2020,Aubert2020}, where galaxy void statistics are complementary to the galaxy power spectrum and baryonic acoustic oscillations \citep[e.g.,][]{Pisani2015,Hamaus2016,Nadathur2019}. Recently, void WL profiles have also been shown to be a powerful cosmological probe \citep[see, e.g.,][]{Melchior2014,Clampitt2015,Cai2015,Barreira2015,Gruen:2015jhr,Barreira:2016ias,Falck2018,Baker:2018mnu,Fang2019}.

While less explored compared with 3D voids, voids can also be identified in projection, such as in the projected galaxy distribution \citep[e.g.][]{Gruen2015,Barreira:2016ias,Sanchez2017,Cautun2018} or in a weak lensing map \citep[e.g.][]{Davies2018,Coulton2019}. 
Here, we follow the latter approach and define WL voids generally as 2D regions within WL convergence maps that contain low convergence or few to no tracers. In a previous work \citep{Davies2018}, we have shown that the lensing profiles of WL voids identified directly in WL convergence maps can be measured with a larger signal-to-noise ratio (SNR) than those of galaxy voids. This is because WL voids correspond to deeper underdensities projected along the line of sight than galaxy voids, and hence they have larger tangential shear profiles. The higher signal-to-noise ratio from WL voids also means that they are better at distinguishing between cosmological models in terms of the signal-to-noise ratio, such as modified gravity models, than galaxy voids \citep{Davies2019b}. Additionally, compared to other WL statistics, WL voids are less affected by baryonic physics \citep{Coulton2019}. 

In this paper we use the \cosmoslics{} simulation suite \citep{Harnois2019} to identify a particular class of WL voids, the {\it tunnels}, for a range of cosmological parameters. We use this data to train a Gaussian process regression emulator, which, combined with Markov chain Monte Carlo, allows us to generate likelihood contours and provide forecast parameter constraints for an \LSST-like survey. 

The tunnel algorithm we use here is one possible choice of WL void finder. In fact, similar to voids identified in the galaxy distribution \citep[e.g.][]{Colberg2008,Cautun2018,Baker:2019gxo}, there are several void finding methods that have been successfully applied to WL maps. For example, \citet{Davies2020} have carried out a detailed analysis on the impact that varying the WL void definition might have on the resulting WL void statistics. 
They have found that the `tunnel' void finding algorithm offers a great trade off between maximising the observable tangential shear profile SNR and minimising the impact of observational noise on the void statistics. Therefore, we limit our analysis to only tunnels, and we defer a more detailed study comparing the parameter constraining powers of different void finders to a future work. For galaxy voids, studies have shown that combining different void definitions can lead to improved cosmological constraints \citep[e.g.][]{Paillas2019}.

The layout of the paper is as follows. In Section \ref{sec:theory} we outline the relevant theory for WL observations. In Section \ref{sec:Mock data etc} we describe our mock observational data, emulation and likelihood analysis pipeline and void finding algorithm. In Section \ref{sec:statistics} we present the WL void statistics used in our analysis and in Section \ref{sec:parameter constraints forecast} we present our parameter constraint forecasts. Finally we conclude in Section \ref{sec:conclusion}. For completeness, we also have three appendices where we study respectively the accuracy of our emulator, the impact of varying the smoothing scale of WL maps, and present the covariance matrix used in our analysis.

\section{Theory}\label{sec:theory}

The lens equation for a gravitationally lensed image is
\begin{equation}
    \pmb{\alpha} = \pmb{\beta} - \pmb{\theta} \, ,
\end{equation}
where $\pmb{\alpha}$ is the deflection angle between $\pmb{\beta}$, the true position of the source on the sky, and $\pmb{\theta}$, the observed position of the lensed image.
The corresponding Jacobian matrix of the (linear) lens mapping is the deformation matrix \textbf{A},
\begin{equation}
    A_{ij} = \frac{\partial \beta_{i}}{\partial \theta_{j}} = \delta_{ij} - \frac{\partial \alpha_{i}}{\partial \theta_{j}} \, .
    \label{eq:amp mat}
\end{equation}
Under the Born approximation and neglecting lens-lens coupling and other second-order effects, the deflection angle can be expressed as the gradient of a 2D lensing potential $\psi$,
\begin{equation}
    \pmb{\alpha} = \pmb{\nabla}\psi \, ,
    \label{eq:alpha}
\end{equation}
where $\psi$ is given by
\begin{equation}
    \psi(\pmb{\theta},\chi) = \frac{2}{c^2} \int_0^{\chi} \frac{\chi - \chi'}{\chi \chi'} \Phi(\chi' \pmb{\theta},\pmb{\theta}) d\chi' \, .
    \label{eq:lensing potential}
\end{equation}
Here, $\chi$ is the comoving distance from the observer to the source and $\chi'$ is the comoving distance from the observer to the continuously-distributed lenses, which is also the integration variable. $\Phi$ is the 3D lensing potential of the lens, and $c$ the speed of light. $\Phi$ is related to the non-relativistic matter density contrast, $\delta= \rho/\bar{\rho} - 1$, through the Poisson equation 
\begin{equation}
    \nabla^2\Phi = 4 \pi G a^2 \bar{\rho} \delta \, ,
    \label{eq:Poisson equation}
\end{equation}
where $\rho$ is the matter density of the Universe (with a bar denoting the mean), $G$ is the gravitational constant and $a$ is the scale factor. 

Eq.~\eqref{eq:lensing potential} shows that the lensing potential is a line-of-sight integral of the matter distribution from the source to the observer. The contribution that matter at distance $\chi'$ along the line of sight makes to the total lensing potential is weighted by $(\chi - \chi') / \chi \chi'$ and so depends on its distances from the source and observer.

Eq.\eqref{eq:alpha} allows Eq. \eqref{eq:amp mat} to be expressed in terms of $\psi$
\begin{equation}
    A_{ij} = \delta_{ij} - \partial_{i} \partial_{j} \psi \, ,
\end{equation}
where partial derivatives are taken with respect to $\pmb{\theta}$. The  matrix $\pmb{A}$ can be parameterised through the more physically instructive terms convergence, $\kappa$, and shear, $\gamma = \gamma_1 + i\gamma_2$, as
\begin{equation}
    \pmb{A} = 
    \begin{pmatrix}
    1 - \kappa -\gamma_1 & -\gamma_2\\
    -\gamma_2 & 1-\kappa+\gamma_1
    \end{pmatrix}
    \, .
\end{equation}    
This parameterisation allows the convergence and shear to be related to the lensing potential via
\begin{equation}
    \kappa \equiv \frac{1}{2} \nabla^2_{\pmb{\theta}} \psi \, ,
    \label{eq:convergence}
\end{equation}
and
\begin{equation}
    \gamma_1 \equiv \frac{1}{2}\left(\nabla_{\pmb{\theta}_1}\nabla_{\pmb{\theta}_1}-\nabla_{\pmb{\theta}_2}\nabla_{\pmb{\theta}_2}\right)\psi, 
    \quad\quad\quad
    \gamma_2 \equiv \nabla_{\pmb{\theta}_1}\nabla_{\pmb{\theta}_2}\psi,
    \label{eq:shear}
\end{equation}
where $\nabla_{\pmb{\theta}} \equiv (\chi')^{-1}\nabla$. Eq.~\eqref{eq:convergence} can be interpreted as a 2D Poisson equation, and so by substituting Eq.~\eqref{eq:Poisson equation} and Eq.~\eqref{eq:lensing potential} into Eq.~\eqref{eq:convergence}, the convergence can be expressed in terms of the matter overdensity
\begin{equation}
    \kappa(\pmb{\theta},\chi) = \frac{3H_0^2\Omega_{\rm{m}}}{2c^2}\int_0^{\chi}\frac{\chi - \chi'}{\chi} \chi' \frac{\delta(\chi'\pmb{\theta},\chi')}{a(\chi')} d\chi' \, .
    \label{eq:conv source}
\end{equation}
This shows that the observed WL convergence can be interpreted as the projected density perturbation along the line of sight, weighted by the lensing efficiency factor $(\chi-\chi')\chi'/\chi$. Here, the lensing efficiency is greatest at $\chi' = \chi / 2$, when the lens is halfway between the source and the observer. 

The above derivation assumes a fixed source plane. However, in real WL observations, the source galaxies do not occupy a single plane at a fixed distance from the observer. The observed catalogue of source galaxies has a probability distribution $n(\chi)$ that spans over a range of $\chi$ values, and Eq. \eqref{eq:conv source} must be weighted by this source galaxy distribution in order to obtain $\kappa(\pmb{\theta})$ \citep[see, e.g.,][for details]{Kilbinger2015}
\begin{equation}
    \kappa(\pmb{\theta}) = \int_0^{\chi} n(\chi') \kappa(\pmb{\theta},\chi') d\chi' \, .
\end{equation}

In this work, we measure the $\kappa$ profile, $\kappa(r)$, in and around WL voids. However, as $\kappa(r)$ is not directly observable, it is also useful to relate it to the radial tangential shear profile, $\gamma_t(r)$, through
\begin{equation}
    \gamma_{\rm{t}}(r) = \bar{\kappa}(< r) - \kappa (r)
    \label{eq:gamma_t} \;,
\end{equation}
where
\begin{equation}
    \bar{\kappa}(< r) = \frac{1}{\pi r^2}\int_0^{r} 2 \pi r' \kappa(r') dr'
    \;
\end{equation}
is the mean enclosed convergence within radius $r$. Notice that here and throughout this paper we use $r$ rather than $\theta$ to
represent the 2D projected distance from the void centre. 

WL observations rely on accurately measuring the shapes of galaxies, and cross correlating the shapes of neighbouring galaxies. However, any correlation in shape due to lensing is dominated by the random shapes and orientations of galaxies, which is the leading source of noises in WL observations, referred to as galaxy shape noise (GSN).  
Since the lensing signal is weak by definition, when identifying WL peaks (local maxima in the convergence field $\kappa(\pmb{\theta})$) it is convenient to express the convergence relative to the standard deviation of the corresponding GSN component of the field. This is given by
\begin{equation}
    \nu = \frac{\kappa}{\sigma_{\rm{GSN}}}
    \label{eq:nu}
\end{equation}
where $\sigma_{\rm{GSN}}$ is the standard deviation of the contributions to the signal from galaxy shape noise. $\sigma_{\rm{GSN}}$ can be calculated by generating mock GSN maps and applying any transformations also applied to the convergence maps, such as smoothing. Mock GSN maps are generated by assigning to pixels random convergence values from a Gaussian distribution with standard deviation 
\begin{equation}
        \sigma_{\rm{pix}}^2 = \frac{\sigma_{\rm{int}}^2}{2 \theta_{\rm{pix}}^2 n_{\rm{gal}} }
    \;,
    \label{eq: GSN gaussian}
\end{equation}
where $\theta_{\rm{pix}}$ is the width of each pixel, $\sigma_{\rm{int}}$ is the intrinsic ellipticity dispersion of the source galaxies, and $n_{\rm{gal}}$ is the measured source galaxy number density. In this work we use $\sigma_{\rm{int}} = 0.28$ and $n_{\rm{gal}} = 20 $ arcmin$^{-2}$ as will be discussed in Section \ref{sec:Numerical simulations} .

It will also be useful to compare constraints from Eq. \ref{eq:gamma_t} to the standard shear two-point correlation function, which is given by

\begin{equation}
    \xi_{\pm}(\pmb{\theta}) = \langle \gamma_t \gamma_t \rangle \pm \langle \gamma_{\times} \gamma_{\times} \rangle = \frac{1}{2\pi} \int_0^{\infty} dl l P_{\kappa}(l) J_{0,4}(l\pmb{\theta}) \, ,
\end{equation}

where $\gamma_t = -\mathbb{R}(\gamma e^{-2i\phi})$ (equivalent to Eq. \eqref{eq:gamma_t}, but presented for completeness), $\gamma_{\times} = -\mathbb{I}(\gamma e^{-2i\phi})$, $\phi$ is the polar angle of the separation vector $\pmb{\theta}$ , $J_0$ and $J_4$ are the Bessel functions for $\xi_{+}$ and $\xi_{-}$ respectively, and $l$ is the Fourier mode.

\section{Methodology}
\label{sec:Mock data etc}

In this section we describe the methodology followed in this work, including the simulations, mock lensing data, emulation, likelihood analysis and the weak lensing void (tunnels) finding algorithm.

The goal of this paper is to present the maximum constraining power that can be achieved with WL voids, in order to motivate further development such as theoretical models and dealing with observational systematics, all of which will be studied in a future work. We note that \citep{Harnois2020} present a methodology for using the emulated WL peak abundance to constrain cosmological parameters from the {\sc DES} year 1 data whilst accounting for observational systematics, and that this approach can be generalised to any non-Gaussian statistic, which would be appropriate for future WL void studies.

\subsection{Mock Data}
\label{sec:Numerical simulations}

In this work we use mock WL convergence maps generated from the \cosmoslics{} and SLICS simulation suites \citep{Harnois2015,Harnois2018,Harnois2019}, which we briefly outline in this subsection.

\begin{figure}
    \centering
    \includegraphics[width=\columnwidth]{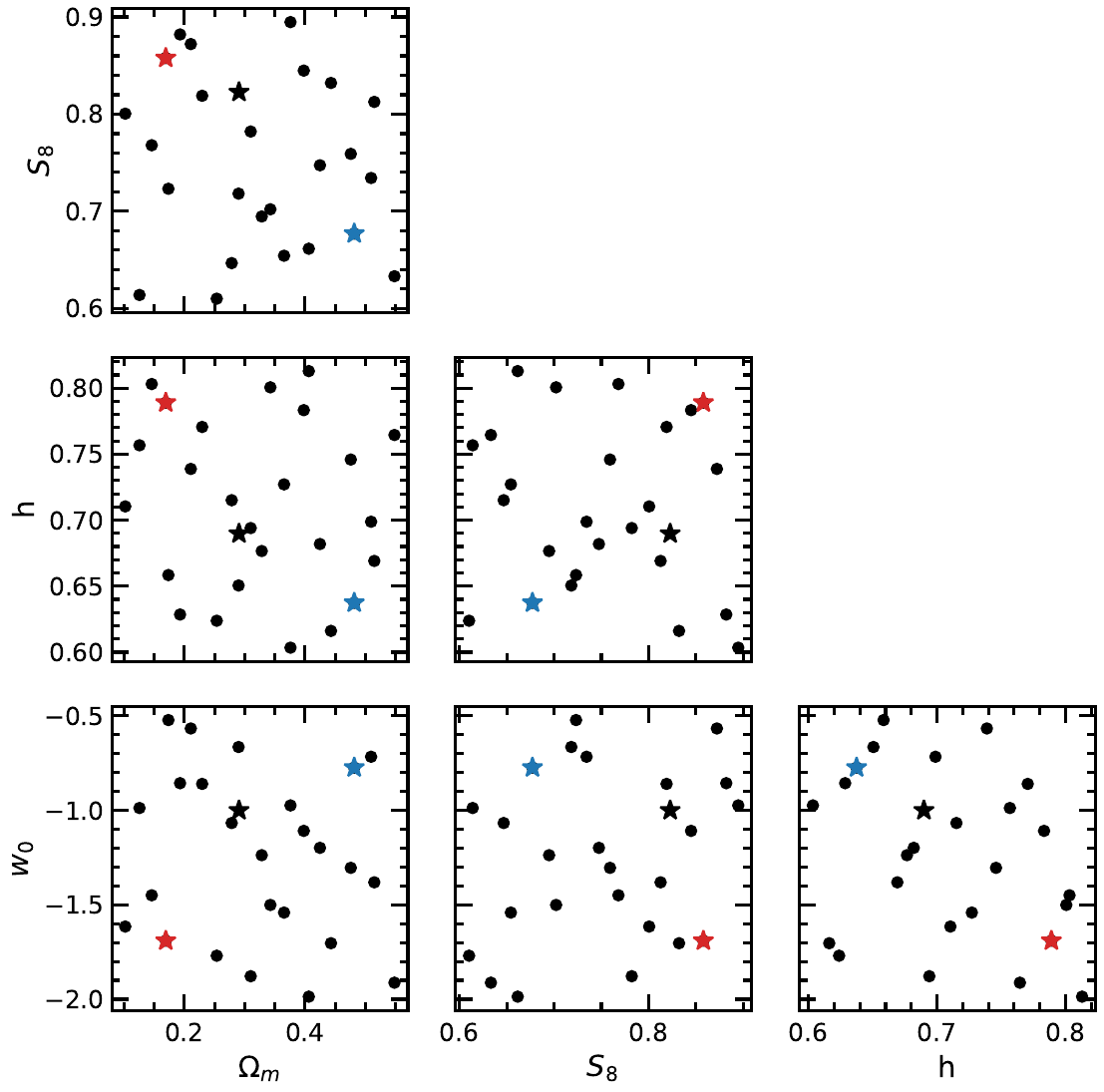}
    \vskip -.1cm
    \caption{The four dimensional parameter space ($[\Omega_m,S_8,h,w_0]$) sampled by the \cosmoslics{} simulation suite. The fiducial cosmology is indicated by a star with parameter values [$0.29$, $0.82$, $0.69$, $-1.00$]. We have highlighted two additional nodes with blue ([$0.48$, $0.68$, $0.64$, $-0.77$]) and red ([$0.17$, $0.86$, $0.79$, $-1.69$]) stars, which are selected as nodes in separate regions of the parameter space, used to exemplify the behavior of WL voids as a function of cosmological parameters.
    }
    \label{fig:cosmo nodes}
\end{figure}

The \cosmoslics{} suite is a set of N-body dark-matter-only simulations run for 26 cosmology nodes in the $[\Omega_m,S_8,h,w_0]$ parameter space. Here $\Omega_m$ is the matter density parameter today, $S_8 = \sigma_8 (\Omega_m/0.3)^{0.5}$, $h = H_0 / 100 {\rm kms}^{-1} \Mpc^{-1}$ is the reduced Hubble constant, and $w_0$ is the dark energy equation of state parameter, which is assumed to be a constant. The $\sigma_8$ parameter is the present-day root-mean-squared matter density perturbation smoothed on $8h^{-1}\Mpc$ scales. 

The four dimensional parameter space is sampled using a Latin hypercube, which is a sampling algorithm designed to give a high interpolation accuracy for a low node count. The exact cosmological parameter space that is modelled by each simulation node is shown in Fig.~\ref{fig:cosmo nodes}. At each node, a carefully-designed pair of simulations are run, for which sampling variance is highly suppressed. This is achieved by selecting a pair of initial conditions out of a large number of random realisations, such that the mean matter power spectrum closely matches the ensemble average. The random phases of this pair of initial conditions are used for all cosmology nodes. The simulation volume is a cube with length $L = 505 \hMpc$, with $N = 1536^{3}$ dark matter particles.

For each node, 50 pseudo-independent light-cones are constructed by resampling projected mass sheets, which are then ray-traced under Born approximation to construct lensing maps and catalogues \citep[see][for full details about the light-cone and catalogue construction]{Harnois2019}. 

We use the \cosmoslics{} source catalogue down-sampled to match $\LSST$ specifications with a source redshift distribution of $z_s = [0.6,1.4]$, which gives a conservative source galaxy number density of $20 \arcmin^{-2}$. From this we generate 50 WL convergence maps for each of the 26 cosmology nodes, with a sky coverage of \map{10} each and pixel grid of dimensions $3600^{2}$ \citep{Giblin2018}. These maps are smoothed with a Gaussian filter with smoothing scale $\theta_s = 1 \arcmin$.

For estimates of the covariance matrices, we use the SLICS suite to produce 615 WL convergence maps at the fiducial cosmology, which match the properties of the \cosmoslics{} maps. However, unlike the \cosmoslics{} maps, the SLICS maps are fully independent which allows us to completely capture the sample variance of the probes studied in this work. Additionally, the larger number of SLICS realisations relative to \cosmoslics{} allows for larger data vectors in the likelihood analysis when measuring and combining probes.

\subsection{Emulation and likelihood analysis}
\label{sec:Emulation}

In this subsection, we outline the procedure used to test the sensitivity of WL void statistics to the cosmological parameters $\Omega_m$, $S_8$, $h$ and $w_0$. 

The first step is to measure the WL void statistics from the 50 convergence maps for each of the 26 cosmo-SLICS cosmologies shown in Fig. \ref{fig:cosmo nodes}. Then, in order to make predictions of the WL void statistics at arbitrary points in the 4D parameter space shown in Fig.~\ref{fig:cosmo nodes}, we use a Gaussian process (GP) regression emulator from scikit-learn \citep{Pedregosa2011} to interpolate the void statistics between nodes. GP regression is a non-parametric Bayesian machine learning algorithm used to make probabilistic predictions that are consistent with the training data \citep[see, e.g.,][for some of its early applications in cosmology]{Habib2007,Schneider2008}. The emulator requires the training data to sample the parameter space sufficiently, and generally the accuracy of the emulator is limited by the availability of training data. The accuracy of the GP emulator trained on \cosmoslics{} was tested extensively and found to yield few per cent accuracy in its predictions of weak lensing two-point correlation functions \citep{Harnois2019}, density split statistics \citep{Burger2020} persistent homology statistics \cite{Heydenreich2020} and aperture mass statistics \citep{Martinet2020}. In this work the average void statistics and their standard errors at each node are used as the training data for the emulator. We present results showing the accuracy of the emulator in Appendix \ref{app:cross_validation}.

Finally, once the emulator has been trained we use Monte Carlo Markov Chain (MCMC) to estimate the posteriors of the parameters for the entire parameter space and produce likelihood contours. We use the emcee python package \citep{Foreman2013} to conduct the MCMC analysis in this work sampling the 4D parameter space as follows.

We employ a Bayesian formalism, in which the likelihood, $P(\pmb{p}|\pmb{d})$, of the set of cosmological parameters $\pmb{p} = [\Omega_m,S_8,h,w_0]$ given a data set $\pmb{d}$, is given, according to Bayes' theorem, by
\begin{equation}
    P(\pmb{p}|\pmb{d}) = \frac{P(\pmb{p}) P(\pmb{d}|\pmb{p})}{P(\pmb{d})} \, ,
\end{equation}
where $P(\pmb{p})$ is the prior, $P(\pmb{d}|\pmb{p})$ is the likelihood of the data conditional on the parameters, and $P(\pmb{d})$ is the normalisation. In our analysis we use flat priors with upper and lower limits respectively for $\Omega_{\rm m}$: [0.10, 0.55], $S_8$: [0.61, 0.89], $h$: [0.60, 0.81],
$w_0$: [-1.99, -0.52]. These priors match the parameter space sampled by the nodes in Fig. \ref{fig:cosmo nodes}.

The log likelihood can be expressed as
\begin{equation}
    \log(P(\pmb{d}|\pmb{p})) = -\frac{1}{2} \left[\pmb{d} - \mu(\pmb{p})\right]C^{-1}\left[\pmb{d} - \mu(\pmb{p})\right] \, ,
    \label{eq:log likelihood}
\end{equation}
where $\mu(\pmb{p})$ is the prediction generated by the emulator for a set of parameters $\pmb{p}$, and $C^{-1}$ is the inverse covariance matrix. In practice we use the emulator's prediction of a statistic at the fiducial cosmology as the data $\pmb{d}$. This choice is for presentation purposes since it ensures that the confidence intervals are always centred on the true values of the cosmological parameters and thus allows for easier comparisons between multiple probes. The likelihood returns a 4D probability distribution that indicates how well different regions of the parameter space describe the input data $\pmb{d}$. Note that in Eq.~\eqref{eq:log likelihood} we have assumed that the covariance matrix does not vary with a change in the cosmological parameters. 

We calculate the covariance matrices from the $615$ WL map realisations from the SLICS suite which match the fiducial cosmology, and divide it by a factor of $180$ in order to rescale the covariance matrix from a $100$ deg$^{2}$ area to the \LSST{} survey area, which we take as $18,000$ deg$^{2}$. The joint covariance matrix for all probes studied in this work is presented appendix \ref{app:covariance}. We also multiply the inverse covariance matrix by a factor $\alpha$, which accounts for the bias that is present when inverting a noisy covariance matrix \citep{Anderson2003,Hartlap2007}, given by:
\begin{equation}
    \alpha = \frac{N - N_{\rm{bin}} - 2}{N - 1} \, ,
\end{equation}
where $N = 615$ is the number of weak lensing maps that have been used to calculate the covariance matrix and $N_{\rm{bin}}$ is the number of bins for which the statistic is computed. We note however that \cite{Sellentin2016} present an alternative approach to robustly account for the uncertainty in the estimated covariance, via a student-$t$ likelihood distribution.

\subsection{The tunnel algorithm}

\begin{figure*}
    \centering
    \includegraphics[width=2\columnwidth]{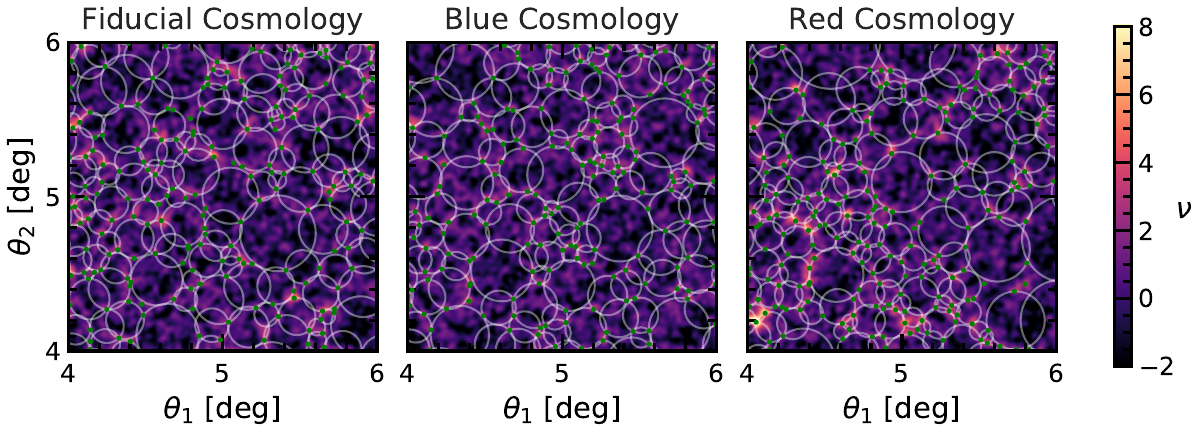}
    \caption{(Colour Online) A visualisation of WL peaks (green points) used to identify the tunnels (white circles) in the WL convergence maps (colour map) for the $\nu > 2$ catalogues. The left panel shows tunnels for the fiducial cosmology, while the middle and right panels show tunnels for the blue and red cosmologies highlighted in Fig.~\ref{fig:cosmo nodes} respectively. The colour bar on the right indicates the convergence field in units of $\nu=\kappa/\sigma_{\rm GSN}$.
    }
    \label{fig:visualisation}
\end{figure*}

To identify WL voids, we use the tunnel algorithm initially proposed in \cite{Cautun2018}, which identifies the largest circles in a 2D tracer catalogue that are empty of tracers. We choose to use this void finding algorithm since, compared with several other common 2D void finders, it gives void lensing profiles with high SNR, whilst also being least affected by the observational noises associated with weak lensing measurements, such as galaxy shape noise \citep{Davies2020}. The tunnel algorithm requires an input tracer catalogue to identify voids. For the identification of WL voids, we use WL peaks as tracers of the underlying convergence field (this avoids the necessity to have a synthetic 2D galaxy map for this analysis). Here we define WL peaks as local maxima in the WL convergence map as in Eq. \eqref{eq:nu}. 

To identify tunnels, we first construct a Delaunay triangulation of the tracers (WL peaks). This produces a unique tessellation of the map with triangles, where each vertex is a tracer and the tessellated triangles enclose no tracers. From each triangle, a corresponding circumcircle can be defined, which is a circle that is directly on top of its Delaunay triangle with all vertices of the latter residing on the circumcircle's circumference. This tessellation is unique, and by definition gives circles that do not enclose any tracers. To avoid identifying the same regions as voids multiple times, we discard any circumcircles whose centers reside inside a larger circumcircle. The resulting list represents our tunnel catalogue, where each tunnel is characterized in terms of the centre and radius of its corresponding circumcircle.

The WL peak catalogues that may be used to identify tunnels contain peaks with a range of amplitudes (or heights) $\nu$. WL peaks of different amplitudes trace different components of the WL map, where the peaks with low or negative amplitudes trace underdense regions of the map, and those with high amplitudes trace overdense regions. Furthermore, peaks with low amplitudes are more susceptible to either being created or contaminated by GSN. It is therefore convenient to generate multiple sub-catalogues of a given WL peak catalogue, by retaining only the peaks with amplitudes larger than a given $\nu$ value. Varying the $\nu$ thresholds allows us to study how the tunnels respond to tracer catalogues with different properties. In this work we use WL peak catalogues with amplitudes of $\nu > 1,2$ and $3$ to identify tunnels, and will also use these $\nu$ values to denote the corresponding tunnel catalogues.

In Fig. \ref{fig:visualisation} we show a visualisation of tunnels identified from catalogues of WL peaks with amplitudes $\nu > 2$. The figure shows WL maps, WL peaks and tunnels for the fiducial cosmology (left), and two sample cosmologies, blue (middle) and red (right) to exemplify the impact of changing cosmological parameters. Here it can be seen in the bottom left part of the panels that the red cosmology, which has the highest $S_8$ value of the three highlighted cosmologies, contains more overdense (orange) regions than the other two cosmologies. The changes in overdensity in the red-cosmology leads to more small tunnels in the bottom left of the panel, and more large tunnels at the top of the panel relative to the other two cosmologies. This highlights how changing the cosmological parameters changes the structure observed in WL maps and the corresponding WL void properties.

\section{Weak lensing void statistics}
\label{sec:statistics}

In this section we present the weak lensing void statistics used in this analysis, showing their abundance in Section \ref{sec:void abundance}, and the tangential shear profiles in Section \ref{sec:lensing profiles}.

\subsection{Void abundance}
\label{sec:void abundance}

\begin{figure*}
    \centering
    \includegraphics[width=2\columnwidth]{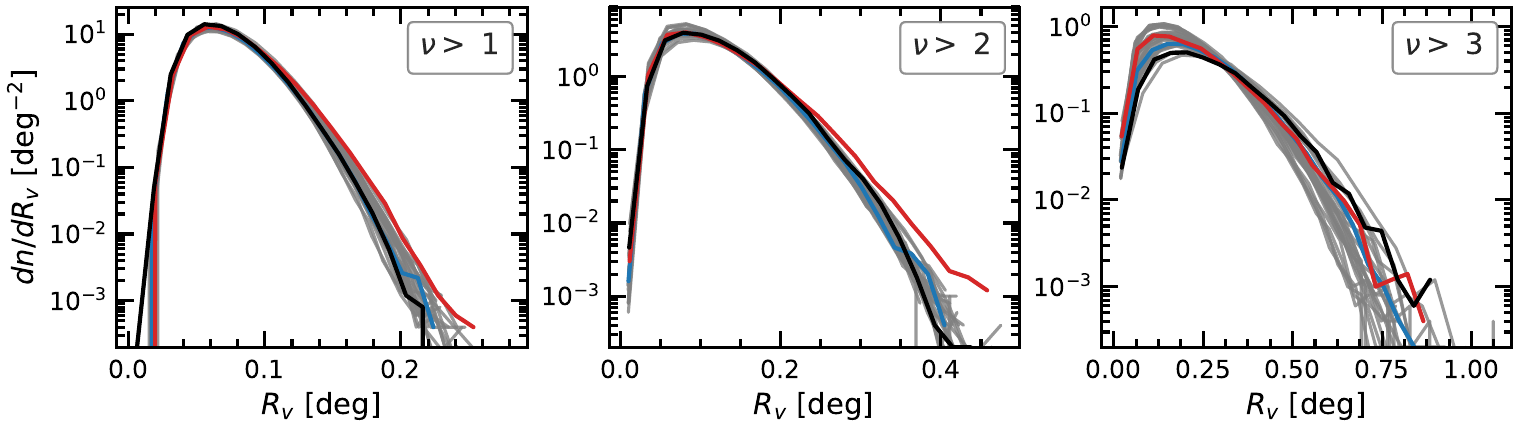}
    \caption{(Colour Online) The differential void abundance as a function of void radius $R_V$. The three panels correspond to voids identified in different WL peak catalogues, with peak heights $\nu > 1,2$ and $3$ (from left to right). The void abundances for all cosmologies in Fig. \ref{fig:cosmo nodes} are plotted in grey. Results for the fiducial (black), red (red) and blue (blue) cosmologies are over-plotted in colour.}
    \label{fig:void abundance}
\end{figure*}

\begin{figure*}
    \centering
    \includegraphics[width=2\columnwidth]{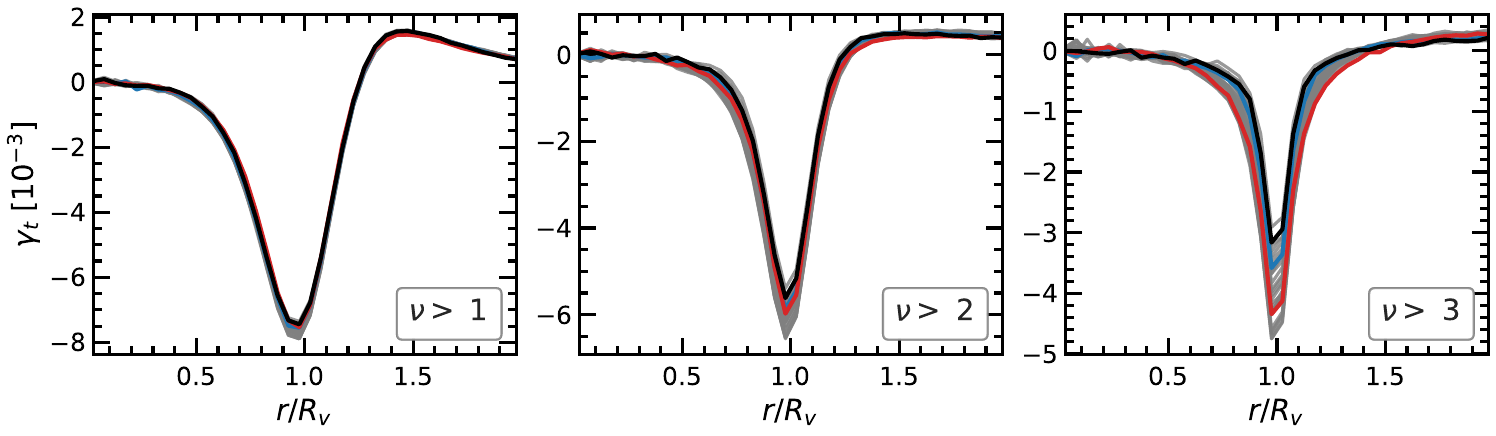}
    \caption{(Colour Online) The tangential shear profiles as a function of re-scaled distance to void centre, $r/R_V$. The three panels correspond to voids identified in three WL peak catalogues with peak heights $\nu > 1,2$ and $3$ (from left to right). The tangential shear profiles for all cosmologies in Fig. \ref{fig:cosmo nodes} are plotted in grey. Results for the fiducial cosmology (black), red (red) and blue (blue) cosmologies are over-plotted in colour. }
    \label{fig:tangential shear}
\end{figure*}

Fig.~\ref{fig:void abundance} shows the differential void abundance per unit area as a function of void radius. The three panels correspond to voids identified in three WL peak catalogues, with peak heights $\nu > 1,2$ and $3$. Void abundances for each of the nodes in Fig. \ref{fig:cosmo nodes} are plotted in grey, the fiducial cosmology in black and two sample cosmologies in colour (blue and red -- corresponding to the two cosmologies in the middle and right panels of Fig.~\ref{fig:visualisation}). 

The figure shows that as the $\nu$ threshold increases, the total number of WL voids decreases (given by the area under the curves), and the average size of the voids increases. The spread in the void abundances over all cosmologies is largest for the $\nu > 3$ catalogue. However, the data is also noisier in this catalogue, because there are fewer peaks with $\nu>3$ and subsequently fewer tunnels.

The red cosmology produces more large voids for the $\nu > 1$ and $2$ catalogues than the fiducial and blue cosmologies. However, the same behaviour is not seen for the $\nu > 3$ catalogue, which may indicate that the sensitivity of the void abundance to specific cosmological parameters changes as fewer tracers are used to identify WL voids. The red cosmology has the largest $S_8$ and smallest $\Omega_m$ compared to the fiducial and blue cosmologies. Increasing $S_8$ or $\Omega_m$ increases the clustering of matter which leads to a wider range of WL void sizes, as we have seen in Fig.~\ref{fig:visualisation}: this is because the enhanced clustering creates more peaks with $\nu>1$ or $2$ in dense regions, reducing the void sizes there, and at the same time reduces the amplitudes of some low peaks in underdense regions, increasing void sizes there. On the other hand, for the $\nu>3$ catalogue, the peaks are sparser in all three cosmologies (hence voids are larger), and the fact that the red cosmology has more peaks at $\nu>3$ again restricts the sizes of its voids, this time affecting the largest ones.

For $\nu > 1$ the fiducial cosmology produces the fewest large voids compared to the red and blue cosmologies, however for $\nu > 3$ it produces the most large voids. The change in relative behaviour between the fiducial, red and blue cosmologies as the $\nu$ threshold increases, indicates that void abundances measured from different WL peak catalogues contain complementary information to each other. We will see this point more clearly later when looking at the constraints from void abundances.

For the $\nu > 1$ catalogue, it is difficult to distinguish between the blue and fiducial cosmology, despite the two cosmologies occupying distinctly separate regions of the parameter space. This is because the cosmological parameters are degenerate, where different combinations of parameters can produce the same void abundances. The degeneracy between parameters also changes between different catalogues.

\begin{table}
    \centering
    \caption{Forecast of percentage uncertainties obtained from various WL void statistics for an \LSST-like survey. The first block of 4 rows show 68\% CL while the bottom 4 rows show 
    95\% CL. In each block, the results shown in the first three lines are quoted from the tightest contours in each figure in Section \ref{sec:parameter constraints forecast} (see first column for more details). In the last line of each block, `$\gamma$-2PCF' stands for the parameter constraints using the cosmic shear two-point correlation functions for the same maps as used for the cosmic void statistics.
    }
    \begin{tabular}{ l cccc } 
        \hline\hline
        Statistic & $\Omega_m$ & $S_8$ & $h$ & $w_0$ \\
        \hline \\[-.25cm]
        \multicolumn{5}{c}{\bf 68\% confidence limits}
        \\[.1cm]
        
        $dn/dR_v$ (combined) & $1.7\%$ & $0.4\%$ & $2.1\%$ & $3.0\%$ \\
        $\gamma_t$ (combined) & $2.3\%$ & $0.5\%$ & $2.3\%$ & $4.4\%$  \\      
        $dn/dR_v$ and $\gamma_t$ (combined) & $1.5\%$ & $0.3\%$& $1.5\%$ & $2.7\%$ \\
        $\gamma$-2PCF & $1.5\%$ & $0.5\%$ & $1.0\%$ & $3.9\%$ \\ 
        
        \hline\\[-.25cm]
        \multicolumn{5}{c}{\bf 95\% confidence limits}
        \\[.1cm]
        
        $dn/dR_v$ (combined) & $3.4\%$ & $0.8\%$ & $4.0\%$ & $5.9\%$ \\
        $\gamma_t$ (combined) & $4.6\%$ & $0.9\%$ & $4.5\%$ & $8.6\%$ \\
        $dn/dR_v$ and $\gamma_t$ (combined) & $2.9\%$ & $0.7\%$ & $2.9\%$ & $5.3\%$ \\
        $\gamma$-2PCF & $3.1\%$ & $1.1\%$ & $2.1\%$ & $8.0\%$ \\ 
        
        \hline\hline
    \end{tabular}
    \label{tab:percentage errors}
\end{table}

\begin{figure*}
    \centering
    \includegraphics[width=2\columnwidth]{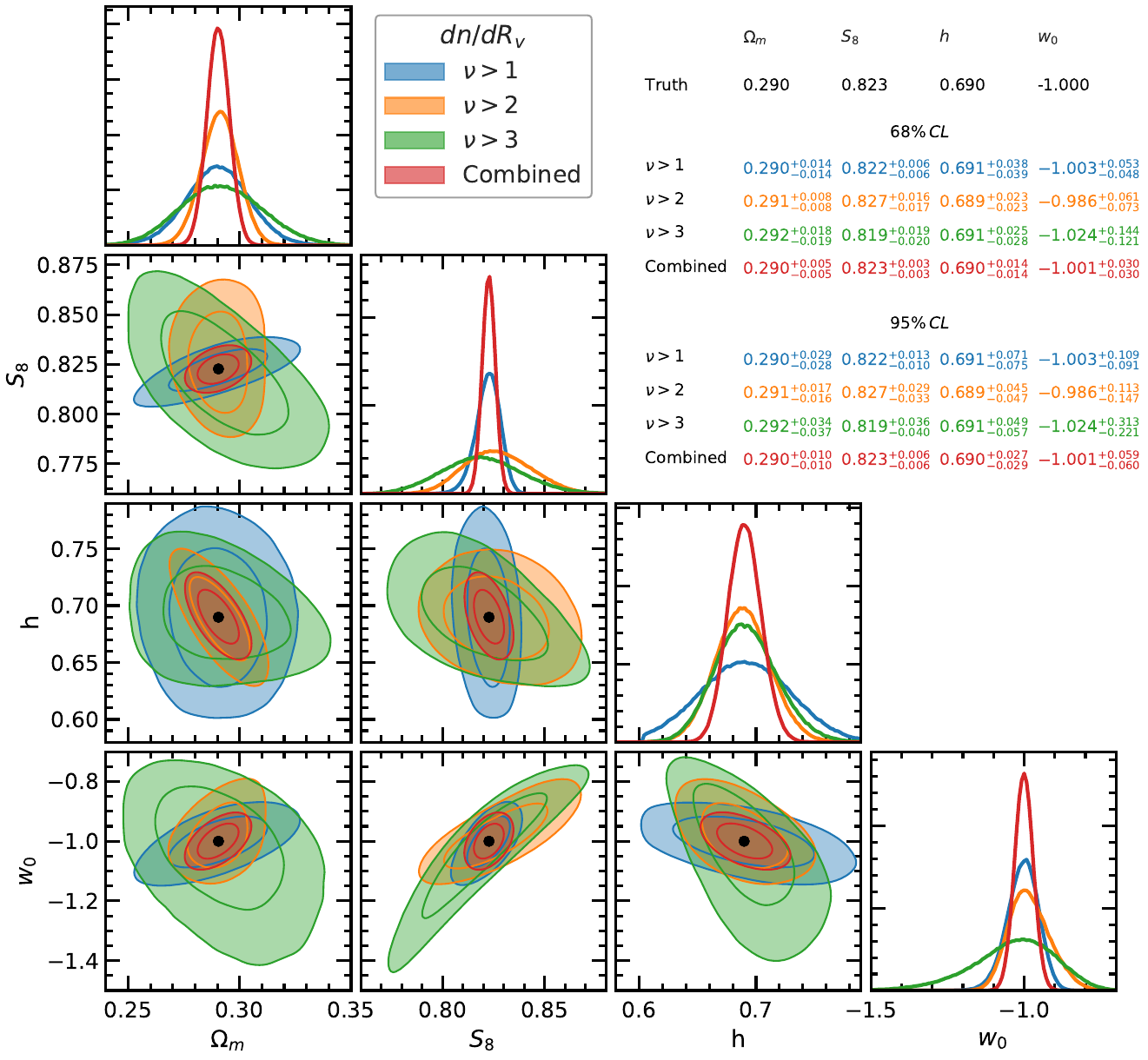}
    \caption{(Colour Online) Constraint forecasts on cosmological parameters measured from void abundances. Contours are shown for WL voids identified in WL peak catalogues with $\nu > 1$ (blue), $\nu > 2$ (orange), $\nu > 3$ (green) and the combination of all three catalogues (red). The true cosmological parameter values used to generate the data are indicated by the black point. The diagonal panels show the 1D marginalised probability distribution, and remaining panels show the marginalised 2D probability contours enclosing the 68\% and 95\% confidence intervals. The table in the top right shows true parameter values (top) and the inferred parameter values for the different peak catalogues with 68\% (upper section) and 95\% (lower section) confidence limits.
    }
    \label{fig:VA contours}
\end{figure*}

\begin{figure*}
    \centering
    \includegraphics[width=2\columnwidth]{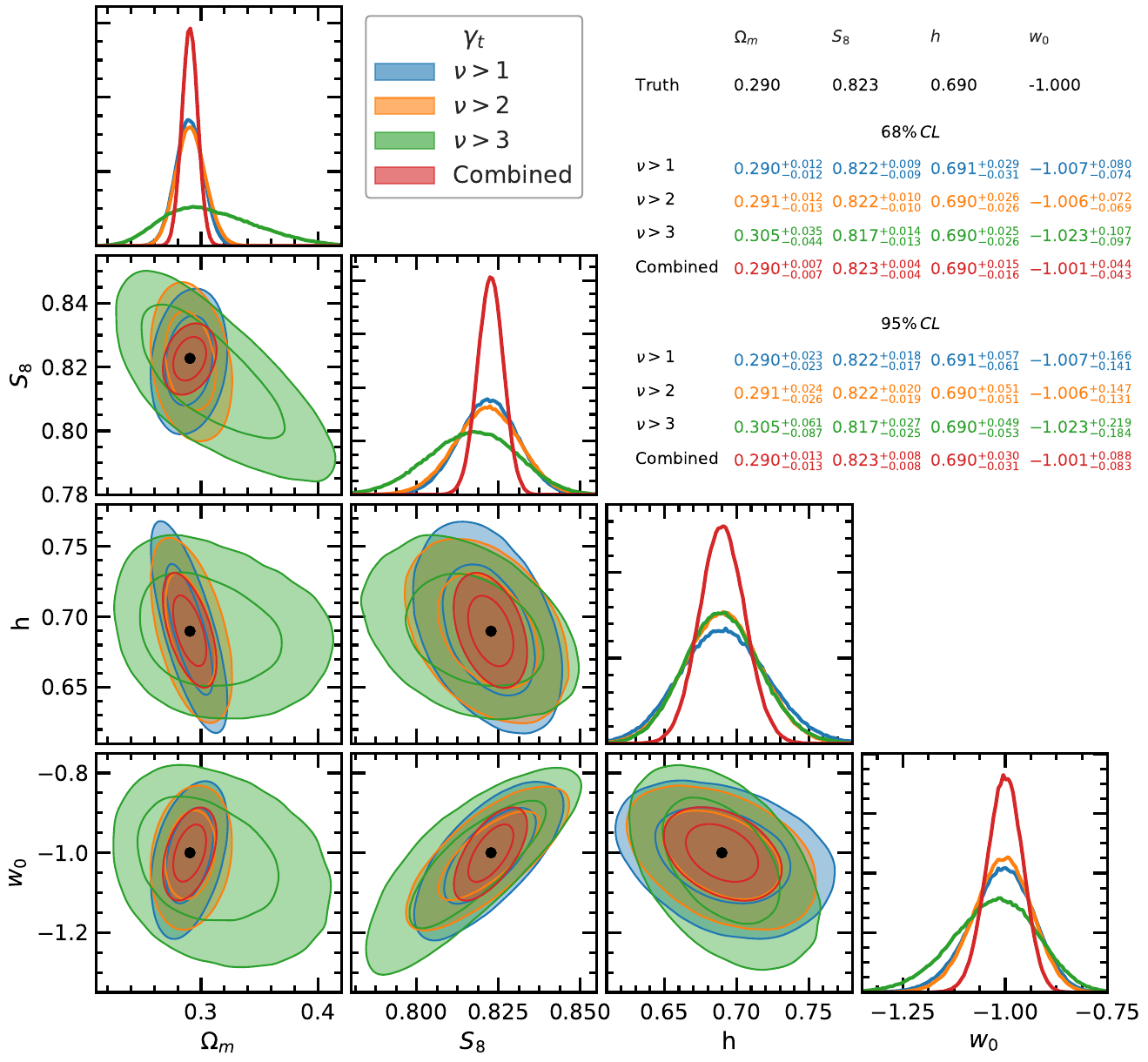}
    \caption{(Colour Online) The same as Fig.~\ref{fig:VA contours} but for the tangential shear profiles. See the caption in Fig.\ref{fig:VA contours} for more details}
    \label{fig:gamma contours}
\end{figure*}

\begin{figure*}
    \centering
    \includegraphics[width=2\columnwidth]{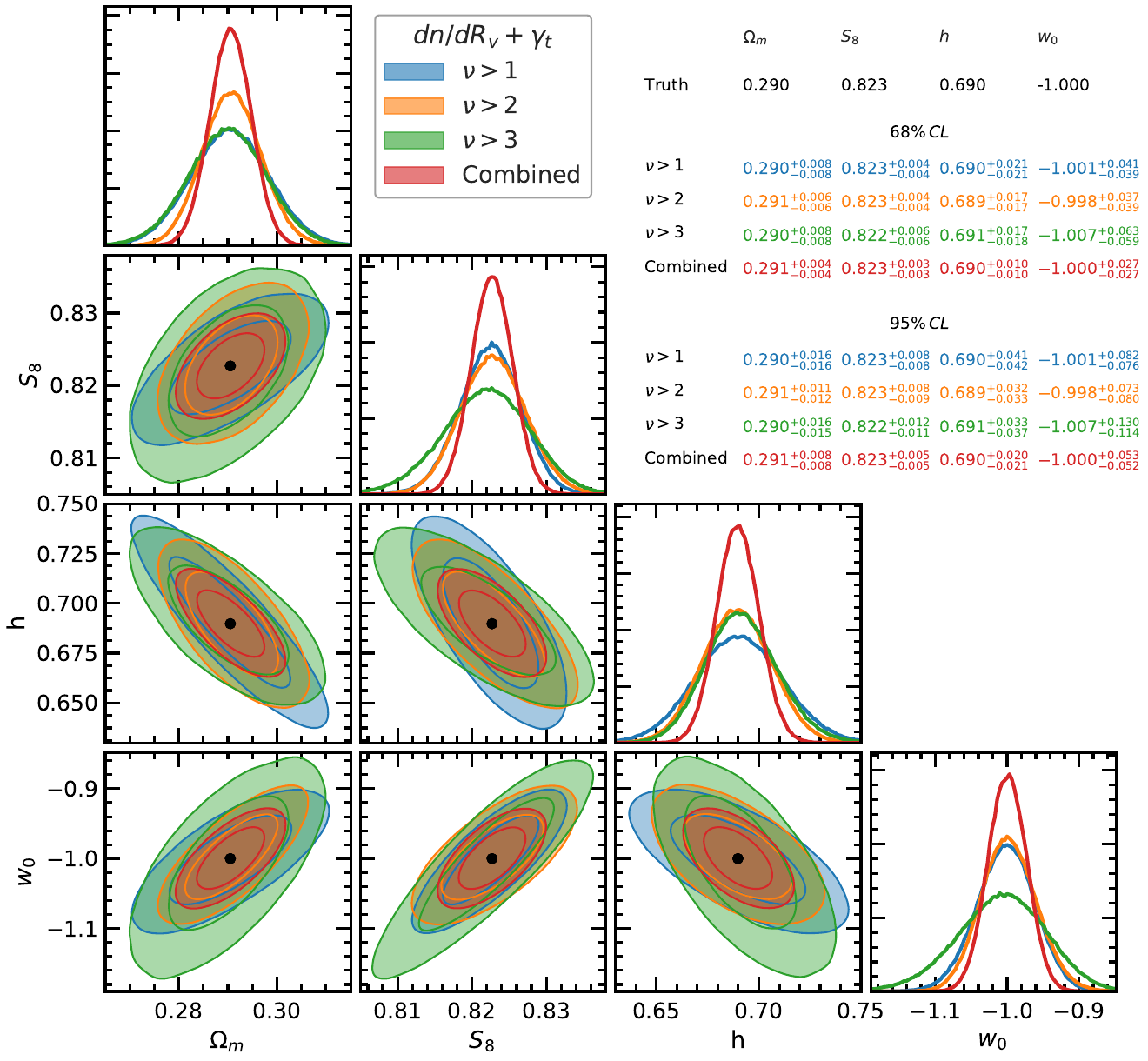}
    \caption{(Colour Online) The same as Fig.~\ref{fig:VA contours} but for the combination of the tangential shear profiles and the void abundance. Results are shown for the three WL Peak catalogues with $\nu > 1$ (blue), $\nu > 2$ (orange), $\nu > 3$ (green). See the caption in Fig.~\ref{fig:VA contours} for more details.}
    \label{fig:combined all nu}
\end{figure*}

\begin{figure*}
    \centering
    \includegraphics[width=2\columnwidth]{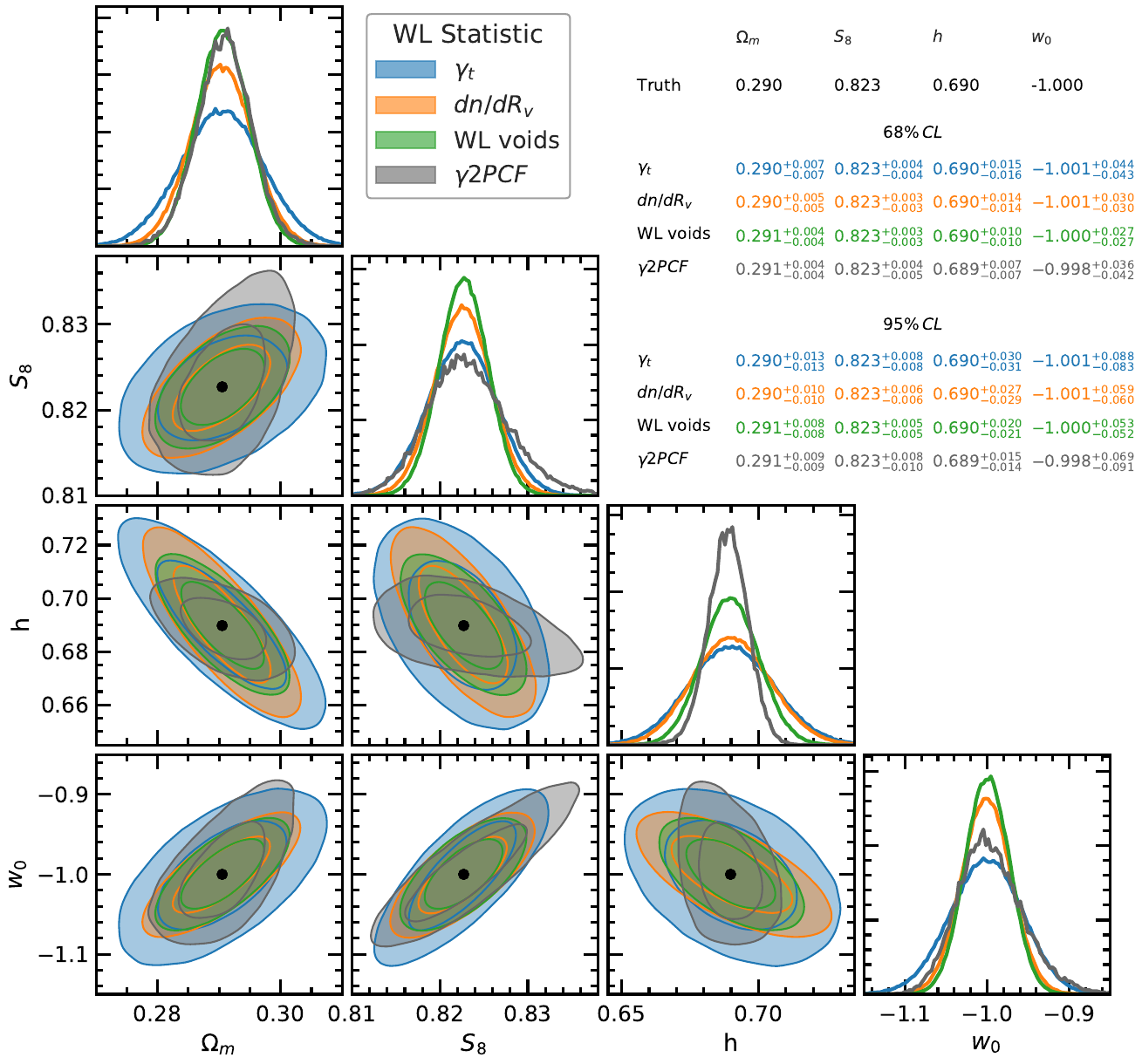}
    \caption{(Colour Online) The same as Fig.~\ref{fig:VA contours} but for the tangential shear profiles (blue) and void abundance (orange). Results are shown for the combination of all three WL peak catalogues. See the caption in Fig.~\ref{fig:VA contours} for more details. Note that, for comparison, we have added the contours from the shear-shear two-point correlation function (without tomography) extracted from the same maps in grey colour, and the corresponding constraints on the parameters are also listed in the table in grey.}
    \label{fig:VA and gamma contours}
\end{figure*}

\subsection{Lensing tangential shear profiles}
\label{sec:lensing profiles}

Fig.~\ref{fig:tangential shear} shows the tangential shear profiles for WL voids, where the panels (from left to right) show WL voids identified in the $\nu > 1,2$ and $3$ catalogues. Tangential shear profiles for all cosmologies are plotted in grey, with the fiducial and two highlighted cosmologies plotted in colour as in Fig.~\ref{fig:void abundance}. The tangential shear profiles are calculated by first measuring the convergence profiles in annuli centered on the void center (pixels are interpolated for small annuli), where the number of annuli used is the lensing profile bin number. The annuli are then stacked as a function of relative angular size ($r/R_V$), weighted by their corresponding void area. Using Eq. \ref{eq:gamma_t}, this is then converted to the tangential shear profiles.

The tangential shear profiles plotted here are negative, which indicates that the WL voids behave like concave lenses and their interiors correspond to underdense regions. The figure shows that as the $\nu$ threshold increases, the depth of the tangential shear profiles at $r/R_V = 1$ decreases, but meanwhile the spread in the amplitude of the tangential shear profiles, as well as the spread in the width of the peak around $r/R_V=1$, amongst all cosmologies, increases. 
Note that the peaks of the tangential shear profiles appear to be narrower as the $\nu$ threshold increases, but this is an artificial consequence of the fact that these plots are made against $r/R_V$ with the void radius $R_V$ larger for larger $\nu$ thresholds.

For the $\nu > 1$ catalogue, the fiducial, red and blue cosmologies all lie on top of each other. For $\nu > 2$, the red cosmology (with the largest $S_8$ value) has a deeper tangential shear profile compared to the other two cosmologies. For $\nu > 3$ the difference in amplitude increases further between the three reference cosmologies, with the fiducial cosmology having the lowest (absolute) amplitude, however the general trend between the three reference cosmologies is the same for all $\nu$ thresholds. Part of this can again be attributed to the high $S_8$ value in the red cosmology, which enhances the clustering of matter, resulting in low-density regions becoming more underdense. However, the three highlighted cosmologies have very distinct values for the other three parameters, in particular $\Omega_m$, which means that an intuitive and yet complete explanation of their relative behaviours is difficult to gauge by eye.

The observation that, although the spread in shapes amongst all cosmologies increases with the $\nu$ threshold, the general order in which they appear does not change, indicates that there may not be much complementary information between tangential shear profiles measured from different $\nu$ thresholds. Also, while the differences between the different cosmologies are larger for the $\nu>3$ catalogue, the fact that there are relatively fewer voids in this catalogue means that its constraining power is not necessarily stronger than the other two catalogues, as we will see shortly.

\section{Parameter constraints forecast}
\label{sec:parameter constraints forecast}

In this section we present parameter constraint forecasts for an \LSST-like survey from the void abundances and tangential shear profiles of WL voids, as well as their combinations.

\subsection{Void abundance constrains}
\label{sec:VA constraints}

Fig.~\ref{fig:VA contours} shows the likelihood contours for measuring the four cosmological parameters with the WL void abundance. The diagonal panels of the figure show the 1D marginalised likelihood distribution and remaining panels show the 2D marginalised likelihood contours. 
For each inference case, the inner and outer contours indicate respectively the 68\% and 95\% confidence limits (CL). As mentioned above, we use the fiducial cosmology as our `observed' data set, which is indicated by the black point. The figure shows results for three $\nu$ thresholds with $\nu > 1$ (blue), $\nu > 2$ (orange) and $\nu > 3$ (green). We also show results for the combination of all three catalogues (red). The table in the top right of the figure indicates the estimated cosmological parameters with their corresponding 68\% (top) and 95\% (bottom) CL, for each of the contours. The tightest contours are for the $\Omega_m - S_8$ plane, which is expected since these are the cosmological parameters to which WL analysis is the most sensitive. 

For nearly every combination of parameters, the three contours for the $\nu > 1$, $2$ and $3$ void catalogues occupy different parts of the marginalised 2D parameter space, or have different degeneracy directions, where most of the overlapping occurs around the true values. As suggested by the behaviour of the three reference cosmologies discussed in Section \ref{sec:void abundance}, this indicates that void abundances measured from different catalogues contain complementary information to each other. We therefore also show parameter likelihood contours for the combination of the WL void abundances from the three catalogues in red. 

We note that for the panels that include $h$, the $\nu > 1$ contours are slightly cut off by the lower prior boundary on $h$. The CLs on $h$ for the $\nu > 1$ catalogue are therefore likely to be slight underestimates compared to the case where a larger prior range on $h$ is used. We do not expand the priors to account for this since the emulator accuracy quickly diminishes outside of the parameter space for which we have training data (which matches our prior range). This does not impact the resulting contours and CLs when all catalogues are combined, since the 'Combined' contours are much smaller and do not approach the prior boundaries.

The `Combined' contours are smaller than any of the individual contours, for all combinations of parameters. This shows that parameter measurements from the WL void abundances are significantly improved when multiple catalogues are used. The $68\%$ and $95\%$ CL percentage accuracy that the combined WL void abundance is able to measure the parameters which is shown in \autoref{tab:percentage errors}.

The WL void abundances for all catalogues are initially measured with 30 bins, which spans the entire range of the WL void sizes measured across all cosmologies. However, since some cosmologies produce more large voids than others. As the void size increases, the point at which the WL void abundance becomes discontinuous due to sample sparsity varies for each cosmology. Therefore, for each catalogue, all bins (for all cosmologies) above the point at which the first discontinuity in any cosmology occurs are discarded. This leads to the WL void abundance being measured with roughly 20 bins, that varies slightly between catalogues, where the largest voids are discarded to due sample sparsity.

Theoretically, the abundance of tunnels identified from a WL peak catalogue depends not only on the number of peaks, but also on their clustering pattern. We therefore expect that the information contained within the WL void abundance and peak correlation functions may have a substantial overlap. The latter probe has been studied in detail by \cite{Davies2019}, with certain scaling properties observed. While it is beyond the scope of the current work, we will conduct a similar analysis by forecasting the parameter constraining power by WL peak two-point correlation functions in a followup study.

\subsection{Tangential shear constraints}

Fig.~\ref{fig:gamma contours} shows likelihood contours for the four cosmological parameters from the tangential shear profiles. The colours of the contours correspond to the same void catalogues as in Fig.~\ref{fig:VA contours}. Again, the contours are smallest in the $\Omega_m$-$S_8$ plane. The figure shows that the contours from $\nu > 1$ and $\nu > 2$ are similar in size, and the $\nu > 3$ contours are significantly larger and in most cases entirely enclose the other contours. All of the contours in this figure, unlike in the case of the void abundances, occupy similar regions of the parameter space, or have similar degeneracy directions. This confirms our conclusion based on the observation of Fig.~\ref{fig:tangential shear}, namely the tangential shear profiles from different peak catalogues do not offer much complementarity.

As in Section \ref{sec:VA constraints}, we combine the tangential shear profiles from all three catalogues to generate `Combined' likelihood contours. Note that for individual catalogues the tangential shear profiles are calculated with 30 bins each. In the likelihood analysis the first two bins are removed. This is because at $r/R_v = 0$, $\gamma_t = 0$, and so the variance is also $0$. This feature induces a singularity close to the origin when inverting the covariance matrix, and so bins near the origin must be removed.

By combining catalogues, we find an improvement in contour size relative the the $\nu > 1$ catalogue, which again suggests that there is complementary information between the different $\nu$ catalogues for the tangential shear profiles. 

The strongest constraints from the tangential shear profiles are for the combined contour. We summarise the 68\% and 95\% CL for the $\gamma_t$ combined case in \autoref{tab:percentage errors}.

\subsection{Constraints by combining void abundance and tangential shear}\label{sec:combined constraints}

In this section we present parameter constraint forecasts for the combination of the WL void abundance and tangential shear profiles. 

Fig. \ref{fig:combined all nu} shows contours for the WL void abundance and tangential shear profiles combined, for the three catalogues $\nu > 1$ (blue), $2$ (orange) and $3$ (green) and for the combination of all three catalogues (red). The smallest contours for an individual catalogue are for the $\nu > 2$ catalogue, and the $\nu > 3$ threshold has the largest contour size, which almost entirely encloses the smaller contours in all cases. This is likely because the number of voids decreases as the $\nu$ threshold increases, meaning that by $\nu > 3$, the statistical uncertainties are large and the constraining power is weakened.

Nevertheless, it is interesting to note that in Fig. \ref{fig:gamma contours} the tangential shear contours for the $\nu > 3$ catalogue are large. The same is also true in Fig. \ref{fig:VA contours} with the WL void abundance for the same catalogue. The resulting contour when the two statistics are combined however is significantly smaller, as shown by the green contour in Fig. \ref{fig:combined all nu}. So even for this catalogue where individual constraints are poor, their combination is highly beneficial.

Fig.~\ref{fig:VA and gamma contours} shows contours for the tangential shear profiles (blue) and WL void abundance (orange) for all three catalogues combined. Note that these contours are also presented as the red contours in Fig.~\ref{fig:VA contours} and Fig.~\ref{fig:gamma contours} respectively. The combination of these two probes, labelled WL voids, is shown by the green contour (repeated from Fig. \ref{fig:combined all nu}).  We also include the shear-shear two-point correlation function constraints as a comparison, which are obtained using the same methodology as that for WL voids. We follow \cite{Asgari2020} and sample the 2PCF using 9 logarithmically-spaced angular separation bins from 0.5 to 300 arcmin, and use both the $\xi_+$ and $\xi_-$ correlation functions, which gives us 18 bins in total. We show the percentage errors (at $68\%$ and $95\%$ CL) for the combination of the shear two-point correlation functions ($\gamma$-2PCF) in \autoref{tab:percentage errors}.

For all combinations of parameters, the WL void abundance contours and the tangential shear contours occupy similar regions of the parameter space and have similar degeneracy directions, where the void abundance contours are slightly smaller than the tangential shear profile contours. Compared to the shear 2PCF, both the WL void abundance and tangential shear profiles are able to constrain $S_8$ with greater accuracy, and the abundance also provides tighter constraints on $w_0$. When all of the WL void statistics are combined, the WL void contours are smaller than the shear 2PCF contours for every combination of parameters, except the $\Omega_m$-$h$ plane, where the two contours have comparable sizes. However, in this plane the two contours also appear to have complementary degeneracy directions. Furthermore, there also appears to be stronger complementary degeneracy directions between the green and grey contours in the $S_8$-$h$ plane and the $h$-$w_0$ plane. Finally, the combined WL void constraints are significantly tighter on $w_0$ compared to the shear 2PCF.

We show the percentage errors (at $68\%$ and $95\%$ CL) for the combination of the WL void abundance and tangential shear profiles over all three catalogues in Table \ref{tab:percentage errors}. The table shows that, compared to the shear 2PCF, the combined WL void statistics are able to provide tighter constraints on $S_8$ and $w_0$ at the $68\%$ CL, and tighter constraints on $\Omega_m$, $S_8$ and $w_0$ at the $95\%$ CL. Although the shear 2PCF provides tighter constraints on $h$, Fig. \ref{fig:VA and gamma contours} shows that the WL void statistics have complimentary degeneracy directions to the shear 2PCF in all panels that include $h$. This indicates that the WL void statistics will also be useful for constraining $h$ when combined with the shear 2PCF.

\section{Discussion and conclusions}
\label{sec:conclusion}

In this paper we have tested the sensitivity of the WL void abundances and tangential shear profiles to four cosmological parameters: $\Omega_m$, $S_8$, $h$ and $w_0$. To this end, we have trained a Gaussian Process emulator with 26 cosmologies sampled in this 4D parameter space using a Latin hypercube, which can be used to predict these two void statistics for arbitrary cosmologies (within the range spanned by the training cosmologies). We have investigated the impact of changing the number of WL peaks used as tracers to identify voids, and ran Markov Chain Monte Carlo samplings from our mock weak lensing data to forecast the accuracy's at which these four parameters can be constrained by a future, \LSST-like, lensing survey, using different combinations of the above WL void statistics.

The results from Fig.~\ref{fig:VA contours} show that the WL void abundance combined over all catalogues gives the tightest parameter constraints, where the greatest sensitivity is to the $S_8$ parameter. This is because the abundances of WL voids identified from WL peak catalogues at different $\nu$ thresholds have different dependencies and degeneracy directions in the studied parameter space. We suspect that there is a close interlink between the void abundance and the peak two-point correlation function, but will defer a detailed study of the latter to a follow-up work. For now, we conclude that complementary information is contained in the abundances of voids from different WL peak catalogues, a fact that should be utilised in order to maximise the use and scientific return of future lensing data.

WL void tangential shear profiles, in contrast, provide slightly less tight constraints on the same cosmological parameters, and the results from different peak catalogues do not seem to be complementary to each other. In particular, for low-$\nu$ peak catalogues such as $\nu>1$ (Fig.~\ref{fig:gamma contours}), there is little degeneracy between $\Omega_m$ and $S_8$; this is because $S_8$ is designed to break the degeneracy between $\Omega_m$ and $\sigma_8$ for standard WL analysis, e.g., shear two-point correlation function, and the low-$\nu$ peaks have little bias with respect to the underlying convergence field, so that their tangential shear profiles follow more closely the parameter dependency of the shear two-point correlation functions. WL void abundances, on the other hand, can have further degeneracy's between $\Omega_m$ and $S_8$ (as seen in Fig.~\ref{fig:VA contours}), indicating that they have different degeneracy directions between $\Omega_m$ and $\sigma_8$ compared with the shear two-point function, and therefore can lead to additional constraints to the latter.

Nevertheless, we highlight that the above conclusions only apply to the 4D parameter space that we have focused on in this work. This may change if additional $\Lambda$CDM parameters such as the spectral index are included. Our results may also be sensitive to changes in curvature, massive neutrinos or other sources of additional physics. In \cite{Davies2019b} we found that the tangential shear profiles are able to distinguish between modified gravity models with a larger signal-to-noise ratio than the void abundance. This suggests that there may be other cosmological parameters not studied here, such as those governing modified gravity laws, to which the tangential shear profile is more sensitive than the WL void abundance. We leave an exploration of this possibility to future works.

Finally, we have found that combining void abundance and tangential shear is another way to obtain tighter parameter constraints. Even for the $\nu>3$ catalogues, for which these two void statistics give poor individual constraints, significant improvement has been found with their synergy.

Overall, we find that weak lensing voids can be a promising cosmological probe to constrain models. The cosmological parameter to which the WL void statistics are most sensitive is $S_8$, which can be measured at the sub percent level (68\% CL). We also find that $\Omega_m$ can be measured to within $\simeq 2\%$, $h$ to within $\simeq 2\%$ and $w_0$ within $\simeq 3\%$ (all 68\% CL). 

As a comparison, we find that parameter constraints from the combination of void abundances and tangential shear profiles are tighter than those from the shear two-point correlation function (which were obtained from the same WL maps, using the same methodology) at the $68\%$ and $95\%$ CLs for all parameters, except $h$. However, the void statistics also have complimentary degeneracy directions to the shear 2PCF for all combinations of parameters that include $h$, which indicates that WL voids are also useful for constraining $h$ when combined with the shear 2PCF, even if the constraints from WL voids alone are not tighter than those from the shear 2PCF.

Additionally, the WL void constraints presented here are for the combination of three peak catalogues. These constraints can be further improved through the inclusion of additional peak catalogues, which may be able to make WL voids a significantly more powerful probe than the shear 2PCF.

We also note that constraints from the shear two-point correlation function can be improved by using tomography \citep{Martinet2020}, and it is therefore also important to test how tomography can improve the constraints from WL void statistics in the future.

Throughout this study, we have adopted a Gaussian smoothing of $\theta_s=1$ arcmin. It may also be interesting to study how the parameter constraints depend on the smoothing scale used to smooth the WL convergence maps. We know that using larger smoothing scales increases the size of the WL voids and reduces their total number \citep{Davies2020}. A larger number of WL map realisations will then be required in order to accurately measure WL void statistics for larger smoothing scales, so we leave such a study to future work. Nevertheless, we have performed a test by using a larger smoothing scale, $\theta_s=2$ arcmin, and in Appendix \ref{app:2 arcmin smoothing} we give a brief summary of the resulting parameter constraints. We can see that the results are similar to what we have found for a 1 arcmin smoothing, cf.~Fig.~\ref{fig:VA and gamma contours}. 

It will also be important to develop an understanding of how the void function is affected by systematics including intrinsic alignments, baryonic feedback, and masking (which can bias statistics measured from convergence maps, e.g. see \cite{Giblin2018} ), which we leave to future study. 

In \cite{Davies2020} we studied the differences in WL void statistics between WL voids identified from different void finders. We found that the tunnel algorithm offered one of the best compromises between high signal-to-noise ratio and small impact from galaxy shape noise in the tangential shear profiles. However, it will also be interesting to assess the constraining power of WL voids identified using other void finders such as the watershed algorithm. The aim is to have a fully comprehensive study of the many different and unexplored ways to use future high-quality weak lensing data to maximise our ability to test cosmological models and constrain cosmological parameters.

\section*{Acknowledgements}
We thank Ludovic Van Waerbeke for the code used in the weak lensing convergence map construction.

CTD is funded by a UK Science and Technology Facilities Council (STFC) PhD studentship through grant ST/R504725/1. MC is supported by the EU Horizon 2020 research and innovation programme under a Marie Sk{\l}odowska-Curie grant agreement 794474 (DancingGalaxies).
JHD acknowledges support from an STFC Ernest Rutherford Fellowship (project reference ST/S004858/1). BG acknowledges the support of the Royal Society through an Enhancement Award (RGF/EA/181006). YC acknowledges the support of the Royal Society through a University Research Fellowship and an Enhancement Award. BL is supported by an ERC Starting Grant, ERC-StG-PUNCA-716532, and additionally supported by the STFC Consolidated Grants [ST/P000541/1, ST/T000244/1].

Computations for the N-body simulations were enabled by Compute Ontario (www.computeontario.ca), Westgrid (www.westgrid.ca) and Compute Canada (www.computecanada.ca). The SLICS numerical simulations can be found at http://slics.roe.ac.uk/, while the cosmo-SLICS can be made available upon request. This work used the DiRAC@Durham facility managed by the Institute for Computational Cosmology on behalf of the STFC DiRAC HPC Facility (www.dirac.ac.uk). The equipment was funded by BEIS capital funding via STFC capital grants ST/K00042X/1, ST/P002293/1, ST/R002371/1 and ST/S002502/1, Durham University and STFC operations grant ST/R000832/1. DiRAC is part of the National e-Infrastructure.

\section{Data Availability}

The data used in this work is available upon request. See \cite{Harnois2019} for more details. 




\bibliographystyle{mnras}
\bibliography{mybib} 



\appendix

\section{Accuracy of the emulator}
\label{app:cross_validation}

\begin{figure*}
    \centering
    \includegraphics[width=2\columnwidth]{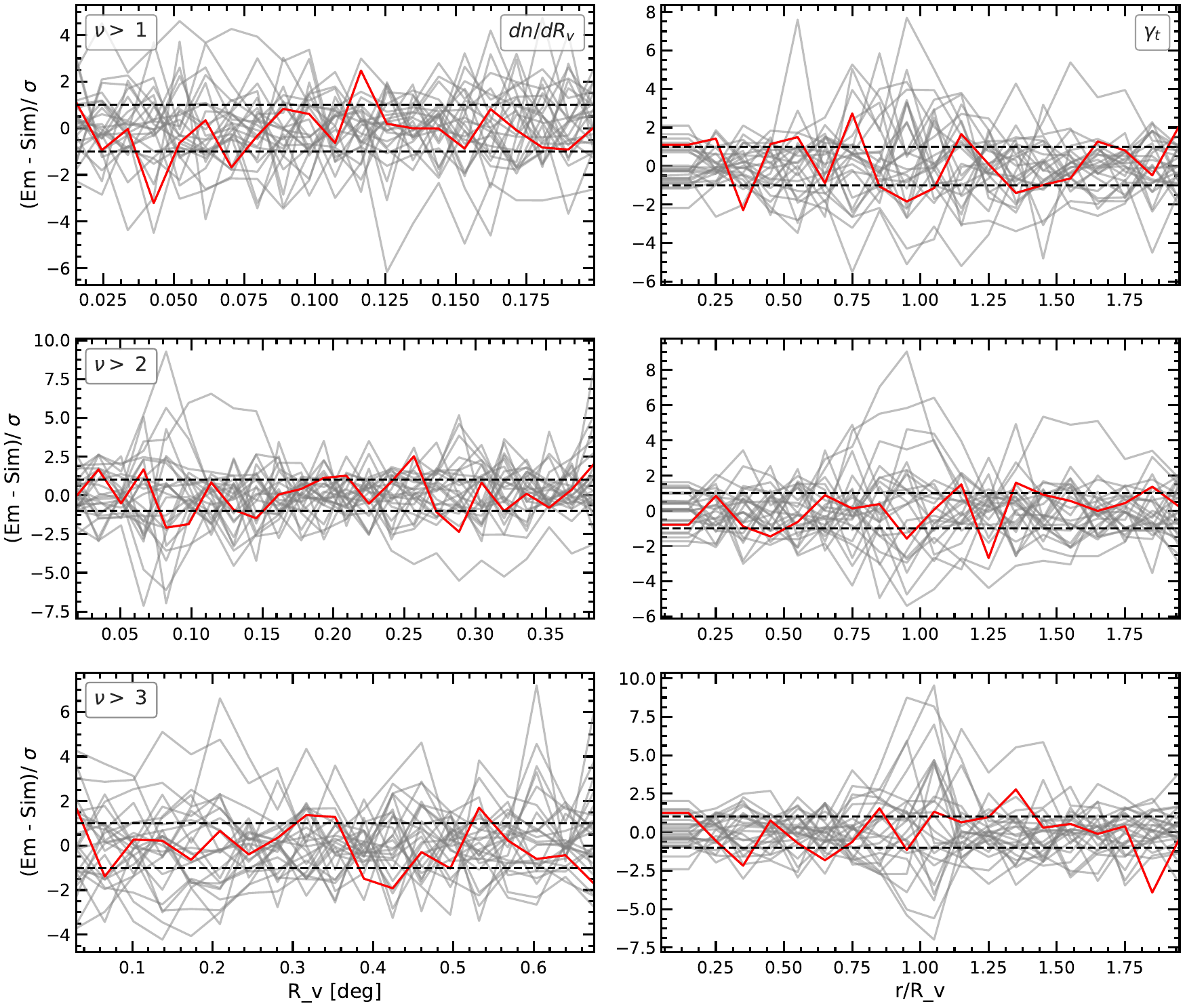}
    \caption{(Colour Online) The cross validation of the emulator accuracy. One node is removed from the training set and the emulation and simulation of the removed nodes are compared relative to its standard error. This is repeated for each of the 26 nodes, which gives an upper limit on the emulator accuracy. The left and right columns show results for the WL void abundance and tangential shear profiles respectively.}
    \label{fig:cross val}
\end{figure*}

In order to test the accuracy of the GP emulator used to interpolate statistics between the cosmological parameter nodes in Fig.~\ref{fig:cosmo nodes}, we perform a cross validation test, which is outlined as follows. First, we remove one node from the training set of simulated data, and train the emulator with the remaining 25 cosmologies. An emulator prediction for the missing node is then calculated. The result is compared to the simulated version, by taking the difference between the two and dividing it by the standard error of the simulated data for that node. The above steps are then repeated 25 more times, by removing a different node from the training set at each iteration, which results in measurements of the emulator accuracy at each node. We note that the above procedure provides an upper limit for the emulator accuracy, since the emulator accuracy increases as more training data is used, and the cross validation measurements uses training data with one less node than the training data used in the main analysis. 

Fig.~\ref{fig:cross val} shows the cross validation test performed for the WL void abundance (left column) and the tangential shear profiles (right column). Results are shown for the catalogues with $\nu > 1,2$ and $3$ in the top, middle and bottom rows respectively. The cross validation test at each node is plotted in grey, with the fiducial cosmology plotted in red. We highlight the fiducial cosmology because we use it as our mock observed data when generating likelihood contours. This makes it the most important region of the parameter space to emulate accurately. 

The figure shows that the emulator accuracy does not vary greatly as a function of the $\nu$ threshold. We find that the emulator is able to accurately predict both the WL void abundance and the tangential shear profiles at roughly the $1\sigma$ level, as denoted by the black dashed lines. 

Regions towards the center of the 4D parameter space will be emulated more accurately than those at the boundary, since there is less training data for the GP emulator to train from at the edges of the parameter space. This is what creates the large spread amongst the grey curves in each panel, where curves towards the center of the 4D parameter space are more accurate, as shown by the fiducial cosmology. We are currently developing a suite of simulations to sample areas of the \cosmoslics{} parameter space more densely, which will help to further improve the accuracy of the emulator by providing more training cosmologies that more densely sample the parameters space through Latin hypercubes or other node design schemes.

\section{The impact of the map smoothing scale}
\label{app:2 arcmin smoothing}

\begin{figure*}
    \centering
    \includegraphics[width=2\columnwidth]{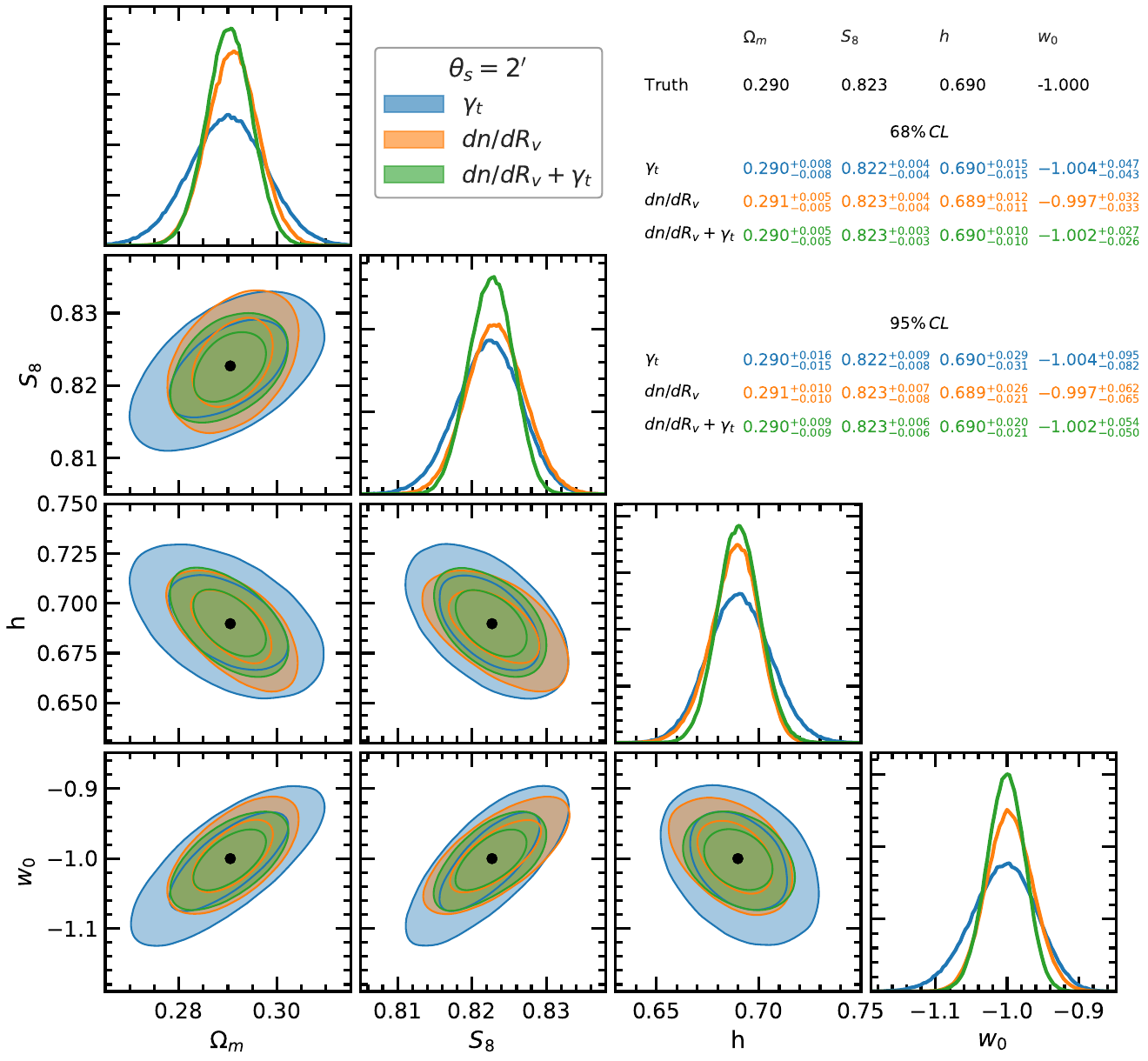}
    \caption{(Colour Online) Likelihood contours for the statistics presented in Table~\ref{tab:percentage errors}, with WL void statistics identified in WL convergence maps smoothed over a $2$ \arcmin{} scale.}
    \label{fig:2 arcmin contours}
\end{figure*}

The analyses carried out in this work used smoothed WL convergence maps, which is required to suppress GSN. However this introduces an additional free parameter in the analysis -- the smoothing scale applied to the maps, where we use a Gaussian smoothing of $1 \arcmin$ in the main body of this work. In \cite{Davies2020} we studied how varying the smoothing scale impacts the resulting WL void statistics, and \cite{J.Liu2015} have shown that parameter constraints from WL peaks can be improved when multiple smoothing scales are combined. It is therefore useful to also show results for a different smoothing scale. 

The likelihood contours for the statistics presented in Table \ref{tab:percentage errors} are shown in Fig.~\ref{fig:2 arcmin contours}, but for a smoothing scale of $2 \arcmin$. These contours behave in a similar way to the case of $1 \arcmin$ smoothing, with tighter constraints coming from the WL void abundance compared to the tangential shear profiles. Overall these constraints are only slightly poorer than for the smaller smoothing scale. 

It is possible to create constraints from combining multiple smoothing scales. However, for brevity, we leave this analysis to a future work. 

\section{Correlation matrix for combined probes}
\label{app:covariance}
In Eq. \eqref{eq:log likelihood} the (inverted) covariance matrix of the data vector is used to calculate the log likelihood. The diagonal elements of the matrix are the variance of each bin in the data vector and the off diagonal elements are the covariance between all possible pairs of bins. When combining multiple probes into a single data vector, it is important to include the cross covariance to ensure that any correlated or duplicate information between the probes is appropriately modelled.

\begin{figure*}
    \centering
    \includegraphics[width=2\columnwidth]{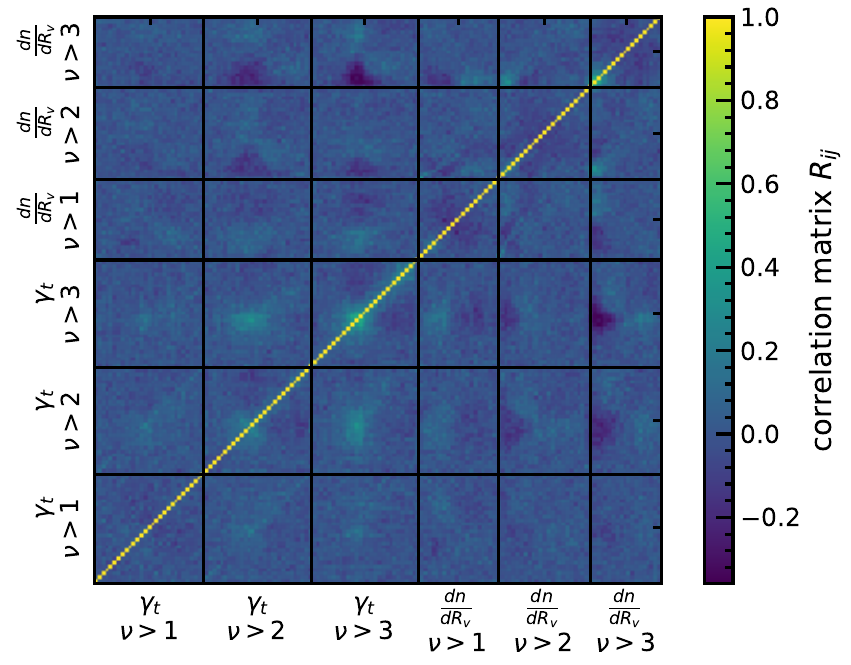}
    \caption{(Colour Online) Correlation matrix for the combination of all WL void statistics presented in this work.}
    \label{fig:correlation}
\end{figure*}

As such, in Fig. \ref{fig:correlation} we present the correlation matrix for the data vector containing each of the WL probes studied in this work, which correspond to the red likelihood contour in Fig. \ref{fig:combined all nu}. The correlation matrix allows for easier visual interpretation and is related to the covariance matrix as follows

\begin{equation}
    R_{ij} = \frac{cov(i,j)}{\sigma_i,\sigma_j}
\end{equation}
Where $R$ is the correlation matrix, $cov$ is the covariance matrix and $\sigma$ is the standard deviation.

\bsp	
\label{lastpage}
\end{document}